\documentclass[12pt]{article}
\usepackage{epsfig}

\parskip=\medskipamount 
plus.3em minus.5em \abovedisplayshortskip=.5em plus.2em minus.4em
\renewcommand{\thesection}{\arabic{section}}

\catcode`@=11

\@addtoreset{equation}{section} \@addtoreset{equation}{subsection}
\def\theequation{\ifnum\value{section}=0 \arabic{equation}\ignorespaces
\else \ifnum\value{section}=-1 A.\arabic{equation}\ignorespaces
\else \ifnum\value{subsection}=0
\thesection.\arabic{equation}\ignorespaces \else
\thesection.\arabic{subsection}.\arabic{equation}\ignorespaces
                             \fi
                        \fi
                   \fi}

{\catcode`\'=\active \def'{{}^\bgroup\prim@s}}

\catcode`@=12

\newcommand{\bq}{\begin{equation}}
\newcommand{\be}{\begin{equation}}
\newcommand{\fq}{\end{equation}}
\newcommand{\ee}{\end{equation}}
\newcommand{\bqr}{\begin{eqnarray}}
\newcommand{\beqs}{\begin{eqnarray}}
\newcommand{\fqr}{\end{eqnarray}}
\newcommand{\eeqs}{\end{eqnarray}}

\newcommand{\rf}[1]{(\ref{#1})}

\def\bop#1{\setbox0=\hbox{$#1M$}\mkern1.5mu
    \vbox{\hrule height0pt depth.04\ht0
    \hbox{\vrule width.04\ht0 height.9\ht0 \kern.9\ht0
    \vrule width.04\ht0}\hrule height.04\ht0}\mkern1.5mu}

\begin{document}
\thispagestyle{empty}

\begin{flushright}
\begin{tabular}{l}
hep-th/0504191 \\
\end{tabular}
\end{flushright}

\vskip .6in
\begin{center}

{\bf  Generalized RSA/DH/ECC via Geometric Cryptosystem}

\vskip .6in

{\bf Gordon Chalmers}
\\[5mm]

{e-mail: gordon@quartz.shango.com}

\vskip .5in minus .2in

{\bf Abstract}

\end{center}

A scheme is presented based on numbers that represent a manifold in 
$d$ dimensions for generalizations of textbook cryptosystems.  The 
interlocking or intersection of geometries, requiring the addition 
of a series of integers $q_j$, can be used to enhance a cryptosystem 
via algorithms based on the form of the geometry.  Further operations 
besides addition of these numbers, e.g. sewing of the geometries, can 
be given such as rotations or contractions. 

\vfill\break

\noindent{\it Introduction}

Textbook RSA, Diffie-Hellman, or elliptic curve cryptostems (ECC) in 
standard representation 
requires the factoring of large numbers into two smaller ones.  The 
time consuming nature of the factoring process protects the security 
of the methods. 

Recently, a polytopic definition of a manifold was introduced in which 
a number specifies a region in space \cite{ChalmersTwo,ChalmersThree}.  
The presence of the number parameterization allows a geometric manipulation 
of the surfaces so that two or more numbers may form a more complicated 
geometry.  The geometry may serve as a algorithm for an encryption process; 
one geometry is specified by a number $p_1$, and a second geometry by 
the number $p_2$, with the two together specifying another geometry.  The 
volumes may specify a surface-dependent encryption process.     

The volume dependence of the combined geometry due to $(p_1,p_2)$, or a 
series of combined geometries $(p_1,\ldots,p_n)$, allows the message sender 
to encode the content in a variety of ways.  Furthermore, the numbers 
may be sent in and manipulated via the secure RSA or DH scheme before 
the geometric gluing of the numbers and subsequent encryption/de-encryption 
using the geometry. 

One adaptation of these protocols allow the user numbers $p_j$ to be chosen 
in an arbitrary manner.  Depending on the number chosen the information 
such as a password may be encrypted in different fashions, which require 
the geometries to unlock.   This is analogous to having one large integer 
$N$ factoring into $pq$ (or $\prod p_j$) for an arbitrary $p$ and $q$; the 
standard methods require the user to possess one number, which is elliptically 
multiplied/calculated.  The presence of the choice of an arbitrary pair 
of numbers in the factoring of a number $N$ results in an exponentiation 
of the possible combinations.

\vskip .2in 
\noindent{\it Polytopes or Polyhedra}

The polytopic surface, or polyhedron, as defined in \cite{ChalmersTwo} 
is described in the number basis.  Take a series of numbers $a_1 a_2 
\ldots a_n$ corresponding to the digits of an integer $p$, with the 
base of the individual number being $2^n$.  In this way, upon reduction 
to base $2$ the digits of the
number span a square with $n+1$ entries.  Each number
$a_j$ parameterizes a column with ones and zeros in it.  The lift of
the numbers could be taken to base $10$ with minor modifications,
by converting the base of $p$ to $10$ (with possible remainder issues
if the number does not 'fit' well).

The individual numbers $a_i$ decompose as $\sum a_i^m 2^m$ with
the components $a_i^m$ being $0$ or $1$.  Then map the individual
number to a point on the plane,

\bqr
{\vec r}_i^m  = a_i^m \times m {\hat e}_1 +
       a_i^m \times i {\hat e}_2 \ ,
\label{latticeentries}
\fqr
with the original number mapping to a set of points on the plane via
all of the entries in $a_1 a_2 \ldots a_m$.  In doing this, a collection
of points on the plane is spanned by the original number $p$, which
could be a base $10$ number.  The breakdown of the number to a set of
points in the plane is represented in figure 1.

A set of further integers $p_j=a_1^{(j)} a_2^{(j)} \ldots a_n^{(j)}$
are used to label a stack of coplanar lattices with the same procedure
to fill in the third dimension.  The spacial filling of the disconnected
polhedron is assembled through the stacking of the base reduced
integers.

The polyhedron is constructed by the single numbers spanning the
multiple layers in 3-d, or by one number with the former grouped as
$p_1 p_2 \ldots p_n$.  The generalization to multiple dimensions is
straightforward.  

The addition of the multiple numbers $a_i^{(j)}$ in each of the 
geometric numbers $q_j$ generate the new geometry and its numbers 
of ${\tilde a}_i^{(j)}$.  The lattice picture is represented in 
\rf{latticeentries}.  

Other operations can be implemented in the sewing of the manifolds.  
There are rotations, contractions, expansions, and displacements of 
the individual geometries, for example.  These operations can be 
implemented before or after the manifolds are molded together.    

\vskip .2in 
\noindent{\it Geometric Manipulation}

The are various ways in which the geometry may be used as an encryption 
method.  A simple one is to take all of the coefficients $a_i^{(j)}$ and 
construct a polynomial, for example, 

\bqr 
P(z)= \sum p_j z^j \ ,  
\label{polynomial}
\fqr
with $p_j=\sum a_i^{(j)} 2^i$, containing the entries of the individual 
rows on the lattice.  This polynomial in \rf{polynomial} is dependent on 
the geometry of the number and could be used as a map to alter information.  
Other polynomials may be found via alternate constructions.

The coefficients associated with the geometry may be used to define an 
L-series, and in turn an elliptic curve.  The coefficients $p_j$ for 
example may be used to count the solutions to a curve 

\bqr  
y^2=x^3 + a x + b \quad {\rm mod ~p} \ , 
\fqr 
with $p$ prime numbers.  The geometry defines an L-series 

\bqr 
\zeta(C,s)=\prod \left(1-a_p p^{-s}+p^{1-2s}\right)^{-1} \ , 
\label{Lseries} 
\fqr 
with the numbers $a_p=p-b_p$, with $b_p$ the number of rational solutions 
to the curve with the modding of $p$.  

Elliptic curves are standard in the RSA and DH schemes, and in alternatives.  
The construction of the geometries, and their gluing together, naturally 
define the elliptic curves.  They could be used in generalizing the 
textbook RSA/DH or elliptic schemes.    

Another method to make a more direct comparison with the previous techniques 
is to have the sender use a number that incorporates with the elliptic 
factorization of these standard methods.  The breakup of a number into 
two smaller ones, i.e. $N=pq$ for a general pair of numbers rather than 
one prefered pair, allows the $p$ and $q$ to be used as the 
individual geometries.  Various numbers $q_j$ could be generated this way; 
these numbers could then be used to enhance the information sharing protocol, 
for both 'password' and message content.

\vskip .2in 
\noindent{\it Concluding Remarks}

The number representation of a multi-dimensional manifold is used to 
provide an enhancement, or alternative, to the well known RSA or 
Diffie-Hellman or elliptic crypto schemes for password encryption.  
Numbers $p_j$ are used to define a geometry, and the sewing of these 
geometries or possible intersection is deduced by adding them.

The geometries are specific to the user, the data, and the receiver and 
dependending on the input manifold a case dependent geometric molding 
is determined.  This is useful in a variety of protected password schemes 
and information sharing. 

\vfill\break


\begin{thebibliography}{99} 

\bibitem{ChalmersOne}
G. Chalmers, {\it Polytopes and Knots}, physics/0503212.

\bibitem{ChalmersTwo} 
G. Chalmers, {\it Integer and Rational Solutions to Polynomial Equations}, 
physics/0503200. 

\bibitem{ChalmersThree}
G. Chalmers, {\it Algebraic and Polytopic Formulation Cohomology}, preprint.

\end{thebibliography}
\end{document}